# Enhanced terahertz transmission through a periodic array of tapered rectangular apertures

**Koijam Monika Devi** *, **Amarendra K. Sarma** and **Gagan Kumar**

Department of Physics, Indian Institute of Technology Guwahati,

Guwahati-781039, India

*koijam@iitg.ernet.in

**Abstract:** We numerically analyse extraordinary terahertz transmission properties of an array of rectangular shaped apertures perforated periodically on a thin metal film. The apertures are tapered at different angles to achieve higher field concentration at the tapered end. The periodic sub-wavelength scale apertures ensure plasmonic behaviour giving rise to the enhanced transmission of a specific frequency mode decided by the periodicity. We compare results of transmission with the rectangular shaped apertures of same parameters and observe a significant increase in the transmission for the tapered case. We have compared results of our numerical simulations with theory and have found them consistent.

**1. Introduction**
THz radiation covers a very narrow range of frequencies (0.3-3 THz) and lies between the microwave and infrared regions of the electromagnetic spectrum. It has been explored for its applications and significance in various fields of science and technology including sensing, imaging, medicine and communication [1-4] . One of the most promising applications of THz radiation is THz imaging as it can penetrate opaque materials such as clothing, paper, cardboard, plastics etc. Also because of the non-ionizing character, they are considered harmless biologically. These features have led to the terahertz imaging applications as a replacement to the X-ray or infrared imaging. Furthermore, terahertz radiations have the ability to provide physical insight of a material that is not accessible using X-ray or infrared radiation [5]. This increases the significance of these radiations.

   THz optics has entered a new phase of research in recent years. A lot of research has been reported to realize the THz devices and improve their performance. In this connection, theoretical investigations on the importance of Spoof plasmons in structured medium for the enhancement of THz transmission have been widely reported [6-8]. Spoof plasmons (SPs) are localized electromagnetic surface wave modes, which can provide an easy way to confine light in sub-wavelength scale structures. Structures supporting SPs include a metal dielectric interface, metal with corrugations or holes, hybrid planar waveguides etc. [9-12] . Using the concept of SPs, interesting phenomenon such as extraordinary transmission has been demonstrated experimentally for an array of apertures fabricated on a metal film in the sub-wavelength regime[13]. Extraordinary transmission is the phenomenon of greatly enhanced transmission for an array of periodic apertures because of the generation of SPs in the periodic structure. Experimental as well as theoretical investigations suggesting the importance of geometry of a structure in the transmission properties has been reported and studied in detail [14-16]. Researchers have examined various shapes and dimensions of the structure including circular, rectangular and square apertures for their terahertz transmission properties. These structures enhance the interaction between the electromagnetic waves and the metal [17, 18]. A 2D array of tapered apertures has been found to efficiently concentrate THz radiation. In 2010, an experimental demonstration by Nguyen et. al reported the transmission properties of a conical tapered aperture fabricated on a metal film and observed an increased concentration of THz radiation at the tapered end. Further enhancement in the concentration was observed by placing periodic apertures in close proximity to the tapered aperture [19, 20].

The ability to focus and concentrate THz radiation would mean a better implementation of it for various applications including near field imaging, biological sensing and development of optoelectronic devices. Inspired with earlier studies, we report here, a new geometry of rectangular tapered apertures patterned in 2-dimensions and study extraordinary terahertz transmission. These structures can be easily fabricated especially in doped silicon making it an experimentally feasible and realizable phenomenon. In this paper, we study the variation of terahertz transmission and concentration by varying the tapering of apertures. The apertures are tapered at different angles and transmission properties are examined in details. As the size of the exit plane is decreased, we calculated a corresponding increase in the transmission efficiency and the concentration magnitude.

## 2. Proposed geometry and simulation

In our numerical analysis, we approximated metal with perfect electrical conductor as it behaves like a perfect conductor at terahertz frequencies. Simulations were performed for an array of rectangular apertures perforated on a thin metal film. We performed simulations for three different sizes of the apertures i.e. length 's' and width 'w' = 500 μm, 323 μm and 232 μm. The thickness (t=500 μm) and the periodicity of the apertures (p=1mm) was kept fixed throughout the study. The arrays were tapered at different angle 'θ' and the transmission properties were studied for each case. Figure 1(a) and 1(b) shows the schematic diagram of the 2D array of tapered apertures (TA). Figure 1(c) shows the side view of a single tapered rectangular aperture, where θ is the angle by which holes are tapered and $D_1$, $D_2$ represents the size of apertures at the input and exit planes. One can calculate $D_2$ for given values of $D_1$ and the tapering angle. For larger values of θ, the size of apertures at the exit plane decreases for fixed width of the substrate.

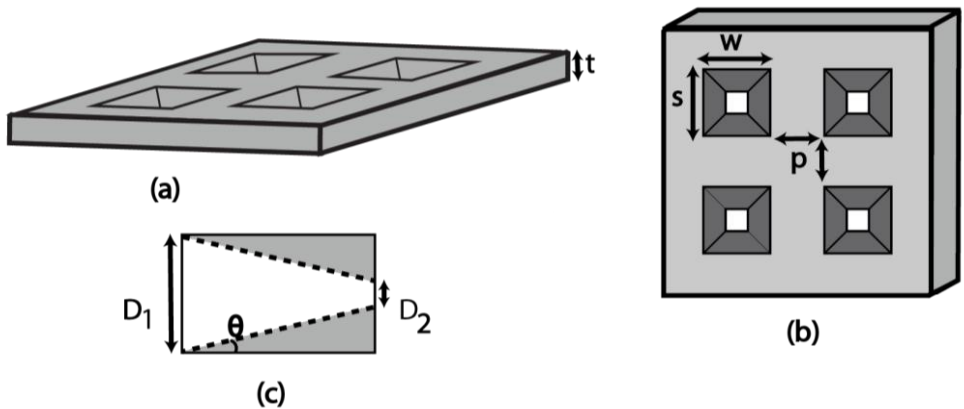

Figure 1: (a) Schematic representation of the tapered apertures arranged periodically in 2-dimension. (b) Front view of the array of tapered apertures. (c) Side view of a single aperture tapered at angle θ w.r.t normal at the input plane.

When the THz radiation is incident on the structure, the periodic nature of the array excites SPs on both interfaces of the structure. Although THz radiation cannot penetrate through metal, the geometry of the structure enables the excited SPs to strongly couple through the apertures. This coupling causes an enhancement in the transmission. This enhancement in transmission can be seen at a particular frequency where the SPs are excited and this frequency is decided by the periodicity of the structure. Moreover, as the THz radiation is incident on the tapered structure, a fraction of the incident radiation gets reflected and a fraction of it gets transmitted. Because of the scattering inside the apertures, the transmitted light gets slowed down and is accumulated at the tapered end causing an enhancement in the concentration of field at the tapered end. The numerical simulations were performed using the technique of finite element method. The whole geometry was meshed with grids of size λ/10 indicating a sub-wavelength regime.

## 3. Results and Discussions

We used frequency domain solver to obtain the transmission properties of the tapered apertures arranged periodically. Figure 2(a) and 2(b) shows the transmission amplitude for an array of tapered apertures in comparison with the reference apertures. In figure 2(a), blue traces shows the transmission amplitude for the tapered apertures with tapering angle $\theta = 10°$, having $D_1$=500 μm, $D_2$= 323 μm, and the red traces represents twice the transmission amplitude for an array of rectangular non-tapered apertures having length 's' and width 'w' both 323 μm which is considered as reference. In figure 2(b), we obtained transmission amplitude for another tapering angle $\theta = 15°$ with $D_1$ unchanged. In this case if one calculates $D_2$ for the same thickness, then it turns out to be 232 μm. The blue and red traces represent the transmission amplitude for tapered and non-tapered array of apertures respectively. In both cases, it is evident that the transmission is greater for the tapered aperture as compared to the non-tapered reference aperture. We calculated more than 20 times enhancement in the transmission amplitude for tapering angle of $15°$ compared with the rectangular apertures.

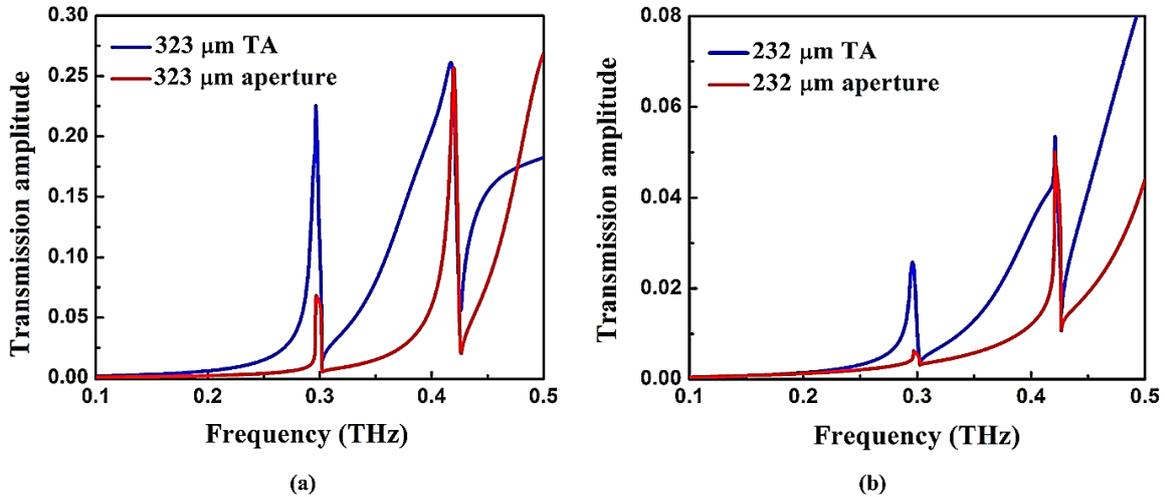

Figure 2: Transmission amplitude for (a) a tapered aperture with parameters: $D_1$=500 μm, $D_2$= 323 μm and $\theta = 10°$. The reference signal corresponds to the rectangular apertures of size 323 μm. (b) Transmission amplitude for periodic tapered apertures corresponding to parameters: $D_1$=500μm, $D_2$= 232 μm and $\theta = 15°$. The reference signal corresponds to the rectangular apertures of size 232 μm.

Figure 3 shows the transmission efficiency of the tapered apertures in comparison with the reference apertures. The transmission efficiency for an aperture is calculated using the formula $t_x^E(\nu) = t_x^E(\nu)/t_{500\mu m}^E(\nu)$, where x is the dimension for the tapered aperture or the reference aperture. The quantity x ranges from 232 μm to 323 μm for different tapering angles. Here, the quantity $t_x^E(\nu)$ is the relative spectral transmission efficiency of any aperture relative to the input. The value $t_{500\mu m}^E(\nu)$ gives the transmission amplitude for an array of rectangular apertures having length and width= 500 μm. Figure 3(a) shows comparison for tapering angle $\theta = 10°$ having $D_1$=500 μm,

$D_2$= 323 μm, and the reference aperture. In the figure, blue traces signify the transmission efficiency for a tapered aperture, while the red one represents the same for reference apertures. Similarly, figure 3(b) shows the transmission efficiency comparison for tapered apertures with tapering angle $\theta = 15°$ having $D_1$=500 μm, $D_2$= 232 μm and the reference aperture. Again the blue and red traces represent transmission efficiency for tapered and non-tapered apertures respectively. One may note that there is a significant enhancement in the efficiency as we increase the tapering angle. In our analysis, we calculated about 25 times enhancement in the transmission efficiency for tapering angle of 15° compared with the rectangular apertures.

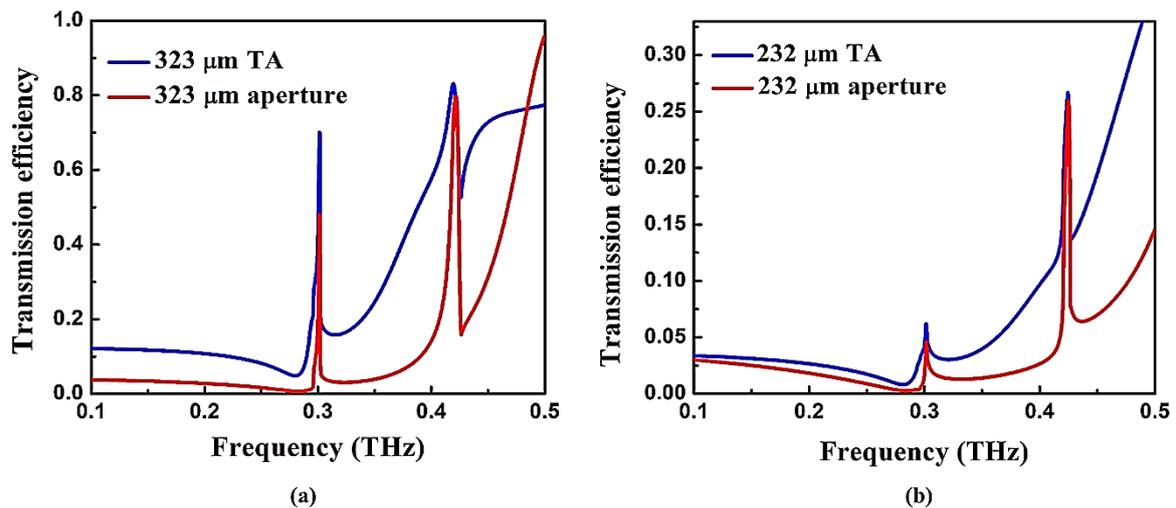

Figure 3: Transmission efficiency for (a) a tapered aperture (TA) with parameters: $D_1$=500 μm, $D_2$= 323 μm and $\theta = 10°$ and the reference aperture, (b) Transmission efficiency for periodic tapered apertures corresponding to parameters: $D_1$=500μm, $D_2$= 232 μm and $\theta = 15°$ and the reference aperture.

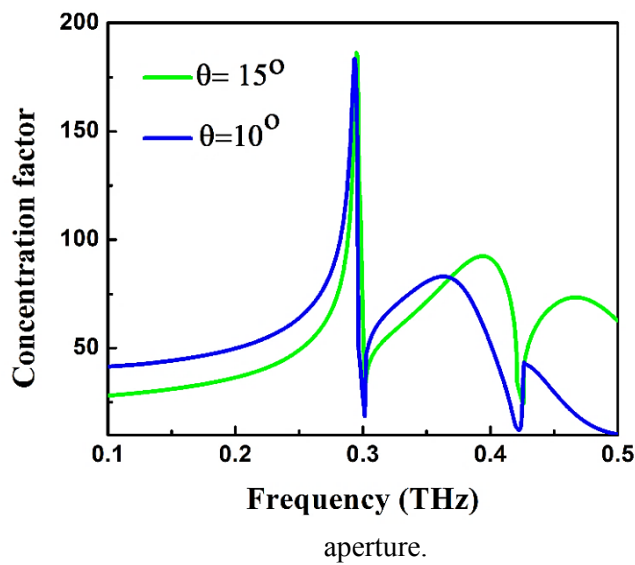

Figure 4: The amplitude concentration factor for tapered apertures for different tapering angle, θ.

Next, we calculate the concentration factor of the terahertz field at the tapered end. The amplitude concentration factor is given by $f_E(v) = t_{x,TA}^E(v)/t_x^E(v)$, where x is the dimension of the aperture. The results are shown in figure 4 for different tapering angles. The blue traces represents concentration factor for the tapering angle of 10° with corresponding dimensions as $D_1$=500 μm and $D_2$= 323 μm and the green traces represents the amplitude concentration factor for tapering angle of 15° with $D_1$=500 μm, $D_2$= 232 μm. It can be noticed that the amplitude concentration factor is greatly enhanced when the tapering angle is increased. In our analysis, we calculated an enhancement of eight times for the tapered aperture with $\theta = 15°$ compared to the tapered aperture having $\theta = 10°$. It is evident from the above results that one can control the transmission efficiency and the concentration factor by changing the angle through which the apertures are tapered. The analysis could be significant for the application where strong terahertz field concentration is required such as near field imaging, sensing etc. It will further help the researcher to design the experiments in this direction.

**4. Conclusion**

In conclusion, we numerically examined the terahertz transmission properties of the periodically arranged apertures tapered at different angles and compare results with the rectangular apertures of same parameters. We observed a significant increase in the transmission amplitude as well as transmission efficiency for the tapered apertures. A subsequent increase in the transmission efficiency as well as concentration of field was noticed as we decreased the size of the aperture at the exit plane. For an apertures tapered at 15°, we calculated a 20 fold increase in transmission and 25 times enhancement in the efficiency compared to the rectangular apertures of same size at the input plane. The terahertz transmission properties can be controlled with the tapering angle as well as the other parameters of the apertures. Our study could be significant for near field imaging technology and in the development of devices operating in the THz domain.

**5. References**
1. Hu, B. and M. Nuss, *Imaging with terahertz waves.* Optics letters, 1995. 20(16): p. 1716-1718.
2. Federici, J.F., et al., *THz imaging and sensing for security applications—explosives, weapons and drugs.* Semiconductor Science and Technology, 2005. 20(7): p. S266.
3. Federici, J. and L. Moeller, *Review of terahertz and subterahertz wireless communications.* Journal of Applied Physics, 2010. 107(11): p. 111101.
4. Krügener, K., et al., *Terahertz meets sculptural and architectural art: Evaluation and conservation of stone objects with T-ray technology.* Scientific reports, 2015. 5.
5. Jepsen, P.U., D.G. Cooke, and M. Koch, *Terahertz spectroscopy and imaging–Modern techniques and applications.* Laser & Photonics Reviews, 2011. 5(1): p. 124-166.
6. Garcia-Vidal, F., L. Martin-Moreno, and J. Pendry, *Surfaces with holes in them: new plasmonic metamaterials.* Journal of optics A: Pure and applied optics, 2005. 7(2): p. S97.
7. Yu, Z., et al., *Terahertz spoof plasmonic coaxial microcavity.* Applied optics, 2014. 53(6): p. 1118-1123.
8. Yin, S., et al., *Spoof surface plasmon polaritons in terahertz transmission through subwavelength hole arrays analyzed by coupled oscillator model.* Scientific reports, 2015. 5.
9. Offerhaus, H., et al., *Creating focused plasmons by noncollinear phasematching on functional gratings.* Nano letters, 2005. 5(11): p. 2144-2148.
10. Pendry, J., L. Martin-Moreno, and F. Garcia-Vidal, *Mimicking surface plasmons with structured surfaces.* Science, 2004. 305(5685): p. 847-848.
11. Maier, S.A., *Plasmonics: fundamentals and applications.* 2007: Springer Science & Business Media.
12. Kumar, G., et al., *Planar plasmonic terahertz waveguides based on periodically corrugated metal films.* New Journal of Physics, 2011. 13(3): p. 033024.
13. Ebbesen, T.W., et al., *Extraordinary optical transmission through sub-wavelength hole arrays.* Nature, 1998. 391(6668): p. 667-669.

14. Koerkamp, K.K., et al., *Strong influence of hole shape on extraordinary transmission through periodic arrays of subwavelength holes.* Physical Review Letters, 2004. 92(18): p. 183901.
15. Ghaemi, H., et al., *Surface plasmons enhance optical transmission through subwavelength holes.* Physical review B, 1998. 58(11): p. 6779.
16. Zhang, X., et al., *Effects of Compound Rectangular Subwavelength Hole Arrays on Enhancing Optical Transmission.* Photonics Journal, IEEE, 2015. 7(1): p. 1-8.
17. Ruan, Z. and M. Qiu, *Enhanced transmission through periodic arrays of subwavelength holes: the role of localized waveguide resonances.* Physical review letters, 2006. 96(23): p. 233901.
18. Lee, J., et al., *Terahertz electromagnetic wave transmission through random arrays of single rectangular holes and slits in thin metallic sheets.* Physical review letters, 2007. 99(13): p. 137401.
19. Diwekar, M., S. Blair, and M. Davis, *Increased light gathering capacity of sub-wavelength conical metallic apertures.* Journal of Nanophotonics, 2010. 4(1): p. 043504-043504-11.
20. Nguyen, T.D., Z.V. Vardeny, and A. Nahata, *Concentration of terahertz radiation through a conically tapered aperture*. Optics express, 2010. 18(24): p. 25441-25448.

**Acknowledgement**

The authors would like to thank the Board of Research in Nuclear Sciences (BRNS), Project no: PHY1P1GK102 for their financial assistance.